\documentclass[prl,twocolumn,aps,a4paper,showpacs,amsfonts,amsmath,amssymb]{revtex4-1}

\usepackage{graphicx}
\usepackage{amsmath} 

\begin{document}

\title{Market Imitation and Win-Stay Lose-Shift strategies emerge as unintended patterns in market direction guesses}
\author{Mario Guti\'errez-Roig} 
\author{Carlota Segura}
\author{Josep Perell\'o}
\affiliation{Departament de F\'{\i}sica Fonamental, Universitat de Barcelona, Barcelona, Spain}

\author{Jordi Duch}
\affiliation{Departament d'Enginyeria Inform\`atica i Matem\`atiques, Universitat Rovira i Virgili, Tarragona, Spain}

\begin{abstract}
Decisions taken in our everyday lives are based on a wide variety of information so it is generally very difficult to assess what are the strategies that guide us. Stock market therefore provides a rich environment to study how people take decision since responding to market uncertainty needs a constant update of these strategies. For this purpose, we run a lab-in-the-field experiment where volunteers are given a controlled set of financial information -based on real data from worldwide financial indices- and they are required to guess whether the market price would go up or down in each situation.
From the data collected we explore basic statistical traits, behavioural biases and emerging strategies. In particular, we detect unintended patterns of behavior through consistent actions which can be interpreted as {\it Market Imitation} and {\it Win-Stay Lose-Shift} emerging strategies, being {\it Market Imitation} the most dominant one. We also observe that these strategies are affected by external factors: the expert advice, the lack of information or an information overload reinforce the use of these intuitive strategies, while the probability to follow them significantly decreases when subjects spends more time to take a decision. The cohort analysis shows that women and children are more prone to use such strategies although their performance is not undermined. Our results are of interest for better handling clients expectations of trading companies, avoiding behavioural anomalies in financial analysts decisions and improving not only the design of markets but also the trading digital interfaces where information is set down. Strategies and behavioural biases observed can also be translated into new agent based modelling or stochastic price dynamics to better understand financial bubbles or the effects of asymmetric risk perception to price drops.
\end{abstract}

%\date{\today}

%\pacs{89.65.Gh, 89.75.Da, 89.65.-s, 05.45.Tp}
\maketitle

\section{Introduction}

Imagine you are a trader. You will be facing the following dilemma everyday: Will the market go up or down? This is the most fundamental question that any financial trader, analyst, advisor, and even non-professional investor with some savings on a given stock is trying to answer using information available from the past -or even from the present-. The key point is to anticipate your action at least one step ahead to what the market will finally do and decide to buy or sell accordingly with the hope of getting some profit from each trade.

Stock price movements are triggered by the matched bids and offers listed in the order book~\cite{campbell:1997}. Bachelier already proposed in 1900 a pure random walk through a binomial process in discrete-time form to describe the resulting price dynamics~\cite{bachelier:1900}. The French mathematician compared the trader as a gambler and admitted in this way that the speculative markets are driven by an important degree of uncertainty. Later on, {\it rational theory} and {\it efficient market hypothesis} better formulated the link between trader's expectations and the evolution of financial prices with the so-called utility function~\cite{samuelson:1937,samuelson:1965}. This robust theory could synthetically be formulated with the following couple of assumptions: (1) information contained in the past is instantly and fully reflected in the current price, and (2) there is no ``free-lunch" without taking any risk which technically means that there is an {\it absence of arbitrage}~\cite{campbell:1997}. The notion of market efficiency brings out a conclusion which appears to be counter-intuitive for a layman: the more efficient the market, the more random is the sequence of price changes, being the most efficient market the one where price changes are completely random and unpredictable~\cite{campbell:1997,farmer:1999}. 

Some studies have found that at least the so-called technical trading strategies are less successful than random strategies~\cite{biondo:2013} and that basic properties in the order book dynamics can be explained by an agent based model which sends to the market buying and selling orders in a complete random way~\cite{farmer:2005}. However, being humans, we still expect to make the correct guess and at least have a better performance than just throwing a coin and this may lead to some behavioural biases~\cite{rabin:1998}. Traders intend to find trends in historical data or hints in any other kind (endogenous and exogenous) of information to reach the inefficiencies of the market which presumably are quickly dissolved in the trading floor~\cite{farmer:1999,lillo:2015}.

One could keep an eye in some financial indexes such as the Dow Jones or the Japanese Nikkei or one could even consider the opinion of a guru. In each case and even dynamically, a trader dives into the ocean of information available and finds out his own recipes and strategies. Several studies have already detected traces of different information explorations in the financial trading activity~\cite{curme:2014,alanyali:2013,preis:2013,bordino:2012,moat:2013,ranco:2015,zhedulev:2014,piskorec:2014,ranco:2016}. Correlations have been found between daily number of mentions of a company in the Financial Times and the daily transaction of that company~\cite{alanyali:2013}, or have quantified possible warning signs of stock market moves based on activity in \textit{Google}~\cite{preis:2013,curme:2014} or {\it Yahoo!}~\cite{bordino:2012,ranco:2016} query volumes, {\it Wikipedia} page views~\cite{moat:2013}, and \textit{Twitter} volume feeds~\cite{ranco:2015}. It is also true that the full amount of information available is neatly impossible to grasp and to analyze in the too limited period of time between transactions. In this sense, it has also been said in the context of human decision making that traders act on the basis of what is been defined as {\it bounded rationality}~\cite{simon:1982,hommes:2013}. Other economists also introduced the {\it prospect theory}~\cite{kahneman:1979,tversky:1986} as an alternative approach to the utility function by considering psychological and framing factors in decision making. Risk perception shifts~\cite{tversky:1981} and judgment under uncertainty biases~\cite{tversky:1974} have been observed in several experiments able to tune the decision frame by for instance changing the formulation of a problem. 

The speculative markets with their rules and mechanisms are indeed a perfect scenario for studying human decision-making mechanisms in an uncertain environment. However, it is difficult, yet not impossible, to monitor the behavior of expert traders~\cite{moro:2009,demartino:2013,frydman:2014,lillo:2015,bouchaud:2016}, and also traders are a biased sample not representative of the society as a whole. For this reason, we designed a controlled experiment that simplistically emulates a trading screen with data from real financial markets, and we asked a large group of volunteers to respond to the question: Based on the information that you have on your screen, do you think that the market will go up or down? The experiment was repeated under three different control settings, and for each of them the volunteers were asked to respond the same question in 25 consecutive rounds. We tracked how many types of information a participant consulted -and for how long- before each decision was made to obtain quantitative measurements for later analysis.

This type of experiment might be understood as a simplified version of the learning-to-forecast laboratory experiments with human subjects where aggregate price fluctuations and individual forecasting behaviour is studied~\cite{hommes:2007,hommes:2013b}. This experimental setting also frames decision making within the so-called Stimulus-Response-Outcome (S-R-O) contingencies~\cite{bland:2012,payzan:2013,platt:2008} and plays with the intrinsic relations between belief and performance \cite{drugowitsch:2014,kovacevic:2015}. In each round, the participant firstly can explore the quantitative information available (that is: historical price in different ways that include moving averages at three different time windows, other markets performance, and expert's advice). Secondly, the participant decides whether the given market will go ``up" or ``down", and thirdly she receives feedback whether her guess was correct or wrong. Using real financial time series from several well-known markets introduces uncertainty, which forces a constant update in participants' strategies due to the extreme variability of the market. This kind of situation where participants have to respond with their guesses receives the name of {\it unexpected uncertainty} or {\it volatility}~\cite{bland:2012,payzan:2013,platt:2008}.

Therefore, we present the results of an experiment with a large heterogeneous group of volunteers aimed to obtain general conclusions concerning human decision making. In particular, the experiment was specifically designed to address to the {\it efficient market hypothesis} and how individuals digest information available~\cite{bouchaud:2016,hommes:2007}. Participants were then solely asked for a binary decision (market ``up" or market ``down"), which constitutes a simple binomial process allowing us to draw conclusions by means of easy-to-apply and quantify statistical tools.

%Among many other aspects, we thus quantitatively explore how a decision depends on the time spent to take this decision and on the amount of information being consulted. We pay attention on the importance of expert's recommendation in player's decision. We show different asymmetries in decision making and how they are related to a risk perception shift~\cite{tversky:1981}. Finally, and most importantly, we closely analyze the emergence in our experiment of the perhaps simplest and most intuitive decision mechanisms developed by humans facing uncertainties: {\it Market Imitation (MI)} and {\it Win-Stay, Lose-Shift (WS-LS)} strategies~\cite{domjan:1986}.

\section{Materials and Methods}

The experiment was carried out inside the context of DAU Festival, a board game fair held in Barcelona during the weekend of 14th and 15th of December 2013. The event was organized by the Institute of Culture of the City Council and attracted 6,000 attendants from Barcelona and its surroundings. The experiment is framed inside the Pop-Up Experiment concept described in~\cite{sagarra:2015}. Participants did not know in advance the details of the experiment and were  asked to play with Mr. Banks (for the participants the experiment was referred as a game) via an interface specifically created and accessible through identical iPads only available in a controlled area -a space with chairs isolated from the rest of the festival-.  At least three researchers simultaneously supervised the experiment at all times, preventing any interaction among the volunteers and avoiding that anybody was repeating the experiment. In order to satisfy privacy issues, all personal data about the participants were anonymized and de-identified in agreement with the Spanish Law for Personal Data Protection and the institutional review board and data protection commissioner of the Universitat de Barcelona. An online informed consent was given by participants for their clinical records to be used in this study.

The data shown in the game was taken from real historical records of different international markets. In particular, we collected 30 series of 25 consecutive days of stock data picked from the period between 01/02/2006-12/29/2009 of daily prices of: the Spanish IBEX, the German DAX and the S\&P500 from United States. The 30 series show different tendencies, specifically 10 with a downwards trend ({\it bearish} market), 10 with upwards trend ({\it bullish} market) and 10 with no specific trend at all ({\it flat} market). Series were assigned randomly to each participant at the beginning of each experiment. Volunteers were told that they were playing with real data but there was no mention about which was the specific market nor about which was the time period of the series to play with. 

\begin{figure}[h]
\centering
\includegraphics[width=0.45\textwidth]{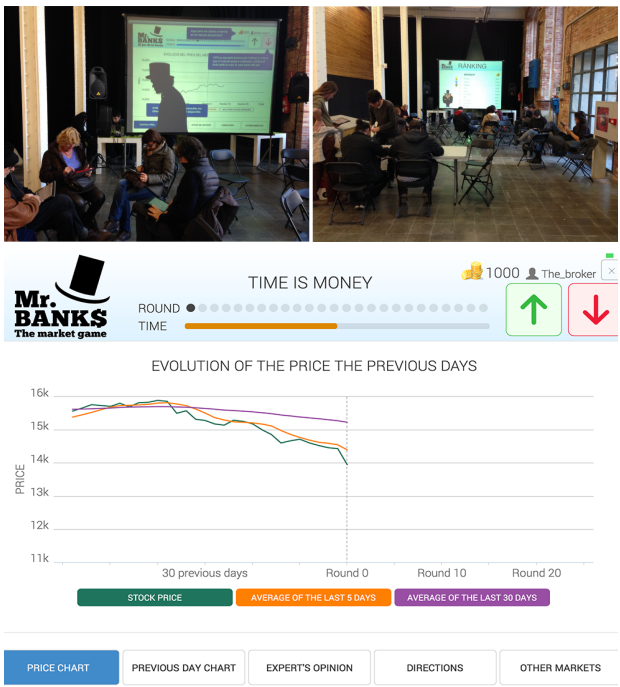}
\caption{{\bf Pictures taken during the experiment and a snapshot of the user interface.} (top) Couple of pictures taken during the experiment. Participants played Mr. Banks game in the experiment delimited area. The general screen showed information about the game and rankings of the best performances in order to capture the attention of attendants in the DAU Festival 2013. (bottom) The screenshots shows the home screen with the buttons to select up (green) and down (red). Participants consulted the different types of information available by clicking the buttons placed below the price chart. The top right corner displays the user alias and the coins being cumulated during the different rounds. See SI for more screenshots of the user interface. }
\label{Figure-1}
\end{figure}

The main setup of the experiment consisted of a screen that showed the information about the market with two buttons to select if the market would go up or down. As shown in top Figure~\ref{Figure-1}, the information available could be easily consulted with 5 buttons that allowed easy navigation across the different screens. The first screen (which corresponds to the home screen) contained the price evolution on a daily basis of the series that should be predicted. The screen was not only plotting the series from the first round, but also the previous 30 days from that round. This screen incorporated two extra buttons that showed a 5-day moving average and/or the 30-day moving average. The second screen showed a chart of the intraday price of the day before. The third screen showed an expert that offered advice using the sentence `{\it Current volatility is high (low) and the price will go ``up" (``down")}' . The expert advice on the price change was correct 60\% of the times but participants were not aware of this. A fourth screen simplified market evolution with just including arrows in green (``up") and red (``down") from market data of the last 30 days. Finally, a fifth screen included information of 9 other indexes (S\&P500, DJI, NASDAQ, FTSE, IBEX, CAC, DAX, NIKKEI, HSI) from 3 different continents with arrows in green (``up") and red (``down") of the last 3 days. Red and green colors were chosen to be consistent with classic colours palette of trading floors infographics (See Supplementary Information, Figs SI.8-SI.12).

The volunteers were asked to guess the price change using this interface 100 times, organized in 4 different scenarios with 25 rounds each. Each guess had a limited time of 30 seconds and, before making a decision, the participant was able to consult the information available according to the scenario constraints. Applying a gamification approach, each participant started the game with 1,000 coins. If the participant made a correct guess, their current number of virtual coins were incremented by 5\% while, if she got a wrong guess, she got a negative return of the same size. This gamifcation approach help us engage the participants into responding the 100 questions of the experiment.

As in a typical experiment in the laboratory, several parameters were tuned in order to know which is the set of conditions of certain phenomena to be produced or what and how it is dependent on. One could easily suspect that time and information are crucial aspects within the making-decision process. In this way, we designed 4 different scenarios in Mr. Banks that could be played by every participant (they were invited to play all scenarios but they could also play just 1 or 2 of them). Participants were randomly assigned to one of the to groups in each scenario once they registered in the experiment. (1) In ``Time is money" all information was available but 50\% of the participants had only 10 seconds to take their decision instead of 30. (2) In ``Information is power", 50\% of the participants had access to all the information available while the others could only consult the price chart of the home screen (without any averages). (3) In ``The computer virus", the available information was limited to only one screen apart from the home screen. For one half of the participants this extra information was randomly chosen while for the other half it could be selected. (4) In ``The trend hunter" 50\% of the participants had access to all the information available and the rest could only see the market directions screen with up and down arrows. They were warned that there was a trend in the financial data without specifying whether it was {\it bearish}, {\it bullish} or a {\it flat} period. Since this fourth scenario was conceptually different from the others we decided to exclude its corresponding data for the latter analysis.

The experiment was also reproduced under the name of ``Hack your Brain" during the course of annual event ``Collective Awareness Platforms for Sustainability and Social Innovation" (CAPS2015, organized by an FP7 European project) that took place in 7 and 8 July 2015 and was hosted in Brussels (Belgium) (See Supplementary Information). A space of 20 square meters at the venue entrance was prepared to carry out the experiment with same protocols and identical interface. 

Same experiment is freely accessible at www.mr-banks.net (in Catalan, Spanish and English). We will shortly update the website to explain the whole project and provide visualization interfaces for all the data gathered. Raw data is available at https://github.com/opensystemsUB/MrBanks.

\section{Results}

283 volunteers were recruited to participate in the experiment in the DAU Festival of Barcelona. The participants took 18,436 valid decisions (89 times they ran out of time) and made 44,703 clicks On the screen. The nature of this Pop-Up Experiment allowed us to study a wider variety of demographics \cite{sagarra:2015}. Thus, from the 283 subjects, 99 (35\%) were females and 184 (65\%) were males. The number of participants by age was distributed as follows: 84 below 15 years old (y.o.), 36 between 16 and 25 y.o., 78 between 26 and 35 y.o., 51 between 36 and 45 y.o., 25 between 46 and 55 y.o., and 9 participants beyond 55 y.o. Additionally they also self-reported about their level of finished studies divided in six groups: None(7), Primary (53), Secondary (37), High School (34), University (148) and Unavailable (4). Participants had no particular expertise in financial markets. We have performed a cohort analysis carefully discussed in the Supplementary Information and in most of the cases we can aggregate the data gathered. 

\subsection{Time, information and expert advice}

\begin{figure*}[t]
\centering
\includegraphics[width=0.85\textwidth]{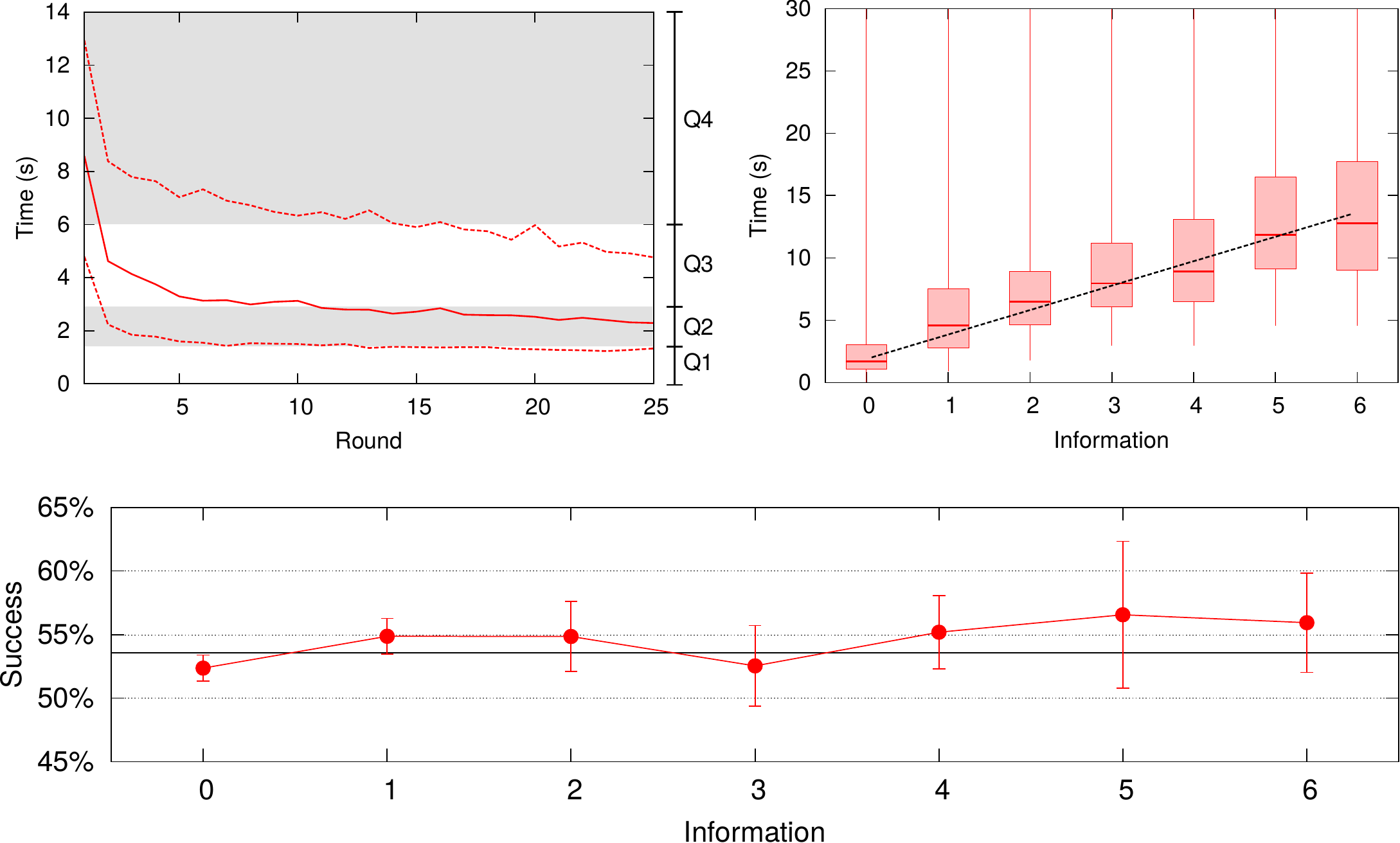}
\caption{{\bf Time, information and performance statistics.} (top left) The time spent in each round by participants rapidly decreases as can be seen with the evolution of the median (solid line) and quartiles (dashed). Shadows within the plot also show the different quartile regions when all the 25 rounds are considered ($Q1=1.614$, $Q2=3.431$, $Q3=6.075$). (top right) Boxplot of the time spent to take a decision as a function of the number of additional pieces information being consulted apart from the price chart displayed at the home screen. Plot shows a linear growth with slope $1.96\pm 0.12$ and origin ordinate $1.89$. (bottom) Success ratio of participants' guesses does not improve when more information is being consulted. Error bars correspond to the standard deviation of a binomial distribution.}
\label{Figure-2}
\end{figure*}

Time, information and expert advice are easy-to-quantify magnitudes to characterize the actions of the participants in our experiment. Time spent in each round is distributed around durations much shorter than the 30 seconds available in the experiment. The fastest quartile of the participant's sample takes their own decision in less than $1.614$ seconds, half part of participants needs at most $3.431$ seconds while the third and slowest quartile of the sample spends more than $6.075$ seconds (see top-left Figure~\ref{Figure-2}). Moreover, such values become quite stable after 5 rounds thus indicating a fast and robust learning curve in contrast with immersive and more sophisticated experiences~\cite{kovacevic:2015}. The average amount (panels) of information being consulted is $2.083 \pm 0.011$ per round. Besides, time spent linearly grows with an slope of $1.96\pm 0.12$ seconds as a function of the number of pieces of information being consulted (see top-right Figure~\ref{Figure-2}). 

The experiment was also designed to measure a well-known behavioural bias in financial markets: the influence of the expert advice~\cite{rabin:1998,tversky:1971,french:1980,engelmann:2009,womack:1996,koestner:1999,andersson:2005}. One of the tabs offered the possibility to consult an expert who was stating whether market will go ``up" or ``down" in the following step. We somewhat arbitrarily fixed that the expert was telling the truth, and thus guessing right, only $60\%$ of the times. Participants thus trusted their opinion with $0.69\pm 0.03$ of probability, which is significantly a higher value (with a $99.87\%$ level of confidence) than the expert forecast reliability. This overreaction phenomena (that can also be understood within the context of the so-called law of small numbers \cite{tversky:1971}) reinforce the mechanisms on how financial analyst's advice is able to generate abnormal price changes~\cite{womack:1996}. Similar phenomena is observed in other situations such as the horse racetrack betting tasks~\cite{koestner:1999}.

The gender cohort analysis in our experiment shows that men consult significantly more information than women (and consequently spend more time to take a decision) while kids consume much less information than adults (and therefore spend much less time to decide). These findings complement recent results studying the effects of endogenous hormones in trading behaviour~\cite{cueva:2015} and also applies in terms of the educational level of the participants (the higher the level, the more time is spent to take a decision) as there exists a correlation between age and educational level. Table SI.4 of the Supplementary Information provides further details of the cohort analysis.

\subsection{Performance, optimistic bias and repetitiveness}

Regarding how well participants can anticipate price change, we obtain 9,879 correct guesses in front of 8,557 incorrect guesses, therefore volunteers had a global success empirical probability of $0.536\pm 0.004$. This result is slightly above of $50\%$ success ratio that one could expect for a fully random process. One could hypothetize that this higher success ratio is achieved by considering all the information provided and using the available time to think about it. However, bottom Figure~\ref{Figure-2} shows that consulting more screens of information, or spending more time looking at the information, do not improve performance -as has also been observed in other contexts such as sports forecasting~\cite{andersson:2005}-. Those participants who played the scenarios where they had information restricted or time shortened (see Materials and Methods) do not show significant differences in their success ratios. Moreover, the fact that success ratio differs in {\it bullish} markets ($0.550 \pm 0.011$), {\it flat} markets ($0.533\pm 0.011$) and {\it bearish} markets ($0.503 \pm 0.011$), lead to think that participants may have a sophisticated behavioural bias behind. The cohort analysis does not show significant differences of success in terms of gender and age although, as mentioned above, there are differences in terms of the time spent to take a decision and the amount of information being consulted.

Table~\ref{table1} shows that the probability that participants choose ``up" ($p(\uparrow)=0.606 \pm 0.004$) is not only very far from a pure random value ($0.5$) but also significantly higher than the empirical probability of the market to go ``up" ($p(\uparrow_M)=0.533\pm 0.004$). Such optimistic bias is also well-known in behavioural finance literature~\cite{barberis:2003} and our experiment allows to make this effect neatly evident. The bias is general among all cohorts studied; age, genre and education level (see Table SI.1 of the Supplementary Information).

It is also known that humans tend to repeat same decision~\cite{rabin:1998}. Table~\ref{table1} shows that this tendency in our experiment has a probability of $0.561\pm0.004$. The value is significantly higher than the $0.5$ from a random behaviour. One could argue that participants believe that market has a trend so they act accordingly. However, such probability is significantly higher than the probability that the market repeats the same outcome ($0.536\pm0.004$) so this behavioural traits can be interpreted as another bias since it relies on a belief perseverance (also called confirmatory bias)~\cite{rabin:1998}. The cohort analysis in Table SI.2 of the Supplementary Information only shows a single exception: the well known fact that children are more inconstant in their decisions~\cite{gutierrez:2014}.

\begin{table*}[t]
\caption{
{\bf Behavioural biases with respect to the market dynamics.} The first column indicates the decision of a single round ``up" or ``down", or with respect to the previous one, ``repeat" or ``change", for either the participants in our experiment (second and third columns) and market data (fourth and fifth). The last to columns compute the difference between humans and market either directly (sixth column) or in terms of Standard Deviation units as defined in the Supplementary Information (seventh column).}
\begin{tabular}{|c|c|c|c|c|c|c|}
\hline
& \multicolumn{2}{|c|}{\bf Subjects} & \multicolumn{2}{|c|}{\bf Market} & \multicolumn{2}{|c|}{\bf Difference}\\
& Decisions & Probability & Ocurrences & Probability & of probablities & in SD units\\
\hline
\hline
``up" & $11137$ & $0.606\pm0.004$ & $10382$ & $0.533\pm0.004$ & $+0.073$ & $+8.37$ \\
\hline
``down" & $7299$ & $0.394\pm0.004$ & $8143$ & $0.467\pm0.004$ & $-0.073$ & $-8.37$ \\
\hline
\hline
``repeat" & $9889$ & $0.561\pm0.004$ & $9445$ & $0.536\pm0.004$ & $+0.025$ & $+4.77$ \\
\hline
``change" & $7732$ & $0.439\pm0.004$ & $8176$ & $0.464\pm0.004$ & $-0.025$ & $-4.77$ \\
\hline
\end{tabular}
\label{table1}
\end{table*}

\subsection{Market Imitation and Win-Stay Lose-Shift emerging strategies}

Our minds have a tendency to introduce biases in processing certain kinds of information. Furthermore, they also create unintended patterns through consistent actions, that is: emerging strategies~\cite{platt:2008,neuringer:1986}. The {\it Market Imitation} (MI), also called {\it automatic imitation} by neuropsicologists~\cite{heyes:2011}, is one possible emerging strategy. It is based on an stimulus response which does not wait for an outcome and just mimics the external input received. In our case this means that the participant's decision has a strong influence of what market did in the previous round. The MI strategy has been carefully studied for instance in rock-paper-scissors game~\cite{cook:2011} and in generation of random sequence by individuals~\cite{jahanshahi:2006}. The S-R-O design of our experiment~\cite{bland:2012} also allows us to study whether the performance in previous round (``correct" or ``wrong" guess) affects participant's decisions. Another possible emerging strategy is the  {\it Win-Stays Lose-Shift} (W-S L-S) that relies on an stimulus response which, in contrast with the MI strategy, considers the outcome of the previous action. In this case, participants repeat their last decision when this was correct, and change the decision when it was wrong guess. This Win-Stays Lose-Shift pattern has also been found in several contexts~\cite{wang:2014,domjan:1986,nowak:1993}. 

To quantify the importance of these two emerging strategies in our experiment, we have computed the mutual information~\cite{cover:1991} (that is: mutual dependence) to measure the influence of the two different emerging strategies in the participant's actions. Mutual information is defined as
\begin{equation}
  \begin{split}
I(X,Y) & = \sum_{x,y} p(x,y) \log \frac{p(x,y)}{p(x)p(y)} \\
& = \sum_{x,y} p(y|x)p(x) \log \frac{p(y|x)}{p(y)},
  \end{split}
\end{equation}
where $X$ and $Y$ are the two random variables. It is defined positive and takes values between 0 and 1, meaning that both random variables are completely independent or that they are perfectly correlated respectively. Mutual information values are given in bits units since we have used the logarithm with base two.

Firstly, in relation to the MI strategy, we compute the mutual information between participant decision series (``up" and ``down") and previous market movements (``up" or ``down"): $0.045\pm 0.010$ bits. And, secondly in relation to the W-S L-S strategy, we compute the mutual information between participant decision series (``up" or ``down") and outcome of previous action (``correct" and ``wrong"): $0.050\pm 0.010$ bits. In both cases, mutual information might seem quite small but it is significantly higher than not only the market self-information case (that is: the mutual information between the series of direction of market changes shifted one day, $0.003\pm 0.010$ bits) but also the participant's self-reflected actions case (that is: the mutual information between the guesses series shifted one step, $0.005\pm 0.010$ bits). Supplementary Information has a specific section discussing all these results.

\begin{figure*}[t]
\centering
\includegraphics[width=0.85\textwidth]{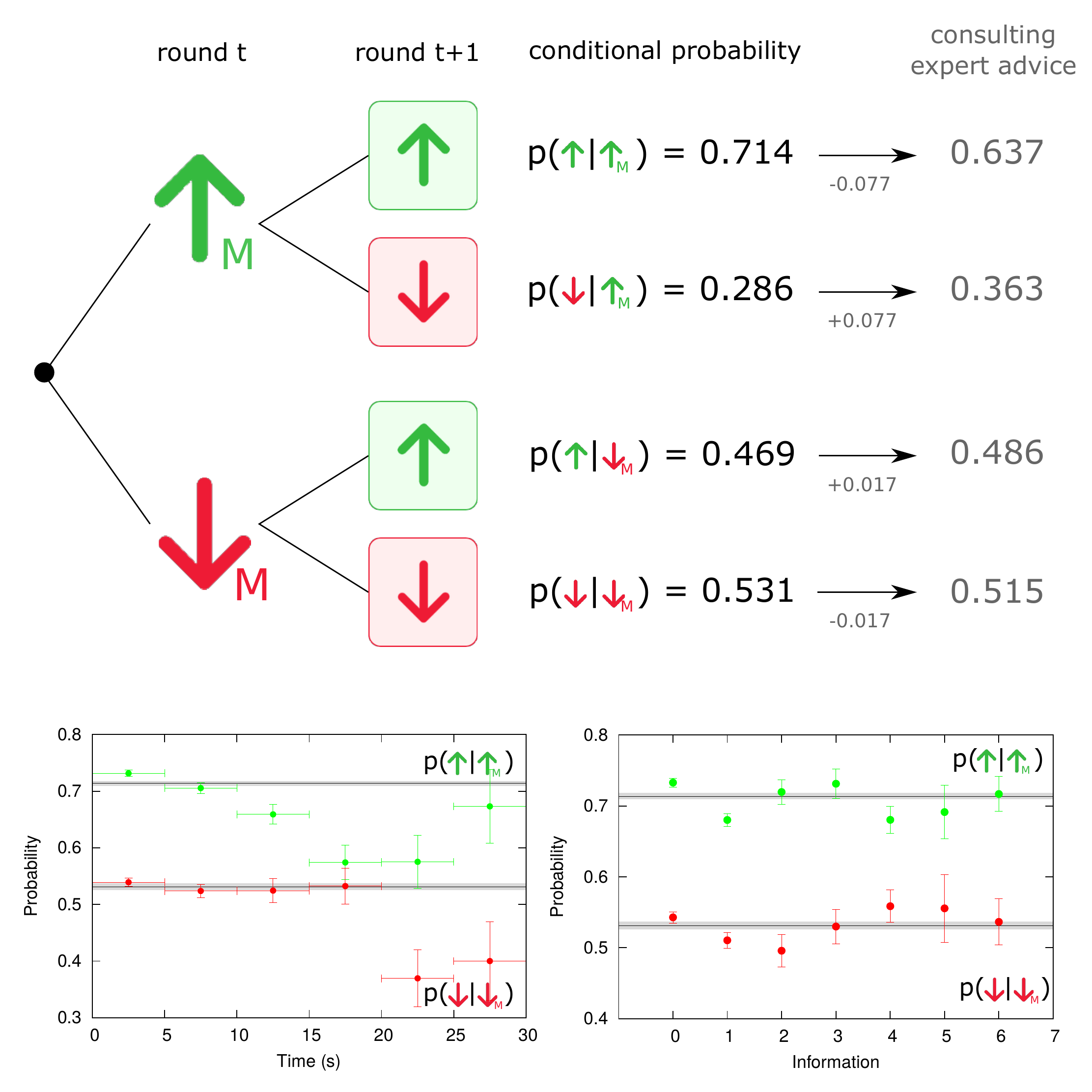}
\caption{\textbf{Decision conditioned to market: the Market Imitation emerging strategy}. Empirical conditional probabilities of guessing whether market will go up or down are positively correlated with market behavior in the previous round. Participants tend to mimic market movement and this behavior is specially important when market went up ($p(\uparrow \vert \uparrow_{\rm M})=0.714\pm 0.005$). The effect is however sensitively diminished when subject consults expert's advice ($0.637$). (down left) The time spent to take a decision plays a significant role when a participant guesses that market will go ``up" (green) in contrast with case when a participant guesses that market will go ``down" (red). (down right) Checking an additional type of information also tends to diminish conditional probabilities but when a participant consults more information, the participant mimics market movement again. Horizontal lines in bottom plots provide aggregated results and shadows their error bars. Error bars represent the standard deviation of a binomial distribution.}
\label{Figure-3}
\end{figure*}

Indeed, we can go one step further by looking at the conditional probabilities related to these two emerging strategies. Figure~\ref{Figure-3} confirms the presence of the MI strategy in our experiment and shows a striking difference between the empirical probability to choose ``up" after market having raised ($p(\uparrow \vert \uparrow_{\rm M})=0.714\pm 0.005$), and the probability to do so but after market having fallen ($p(\uparrow \vert \downarrow_{\rm M})=0.469\pm 0.006$). These two conditional probabilities differ from the unconditional probability to choose ``up" ($p(\uparrow)=0.606\pm 0.004$) by $18.59$ Standard Deviation (SD) units above and $19.04$ SD units below respectively. The imitation is also relevant in the ``down" case. The probability of choosing ``down" conditioned to the market went ``down" is $p(\downarrow \vert \downarrow_{\rm M})=0.531\pm 0.006$ being $20.37$ SD units above the unconditional probability $p(\downarrow)=0.394\pm 0.004$ from Table~\ref{table1}, while $p(\downarrow \vert \uparrow_{\rm M})=0.286\pm 0.005$ is $16.86$ SD units below this value. It is worth mentioning that the MI strategy describes a behavioural bias towards upwards market direction that might be linked to optimistic behaviour and overconfident position with respect positive trends in financial markets~\cite{barberis:2003} and even linked to financial bubbles \cite{jiang:2010,hommes:2013}.

\begin{figure*}[t]
\centering
\includegraphics[width=0.85\textwidth]{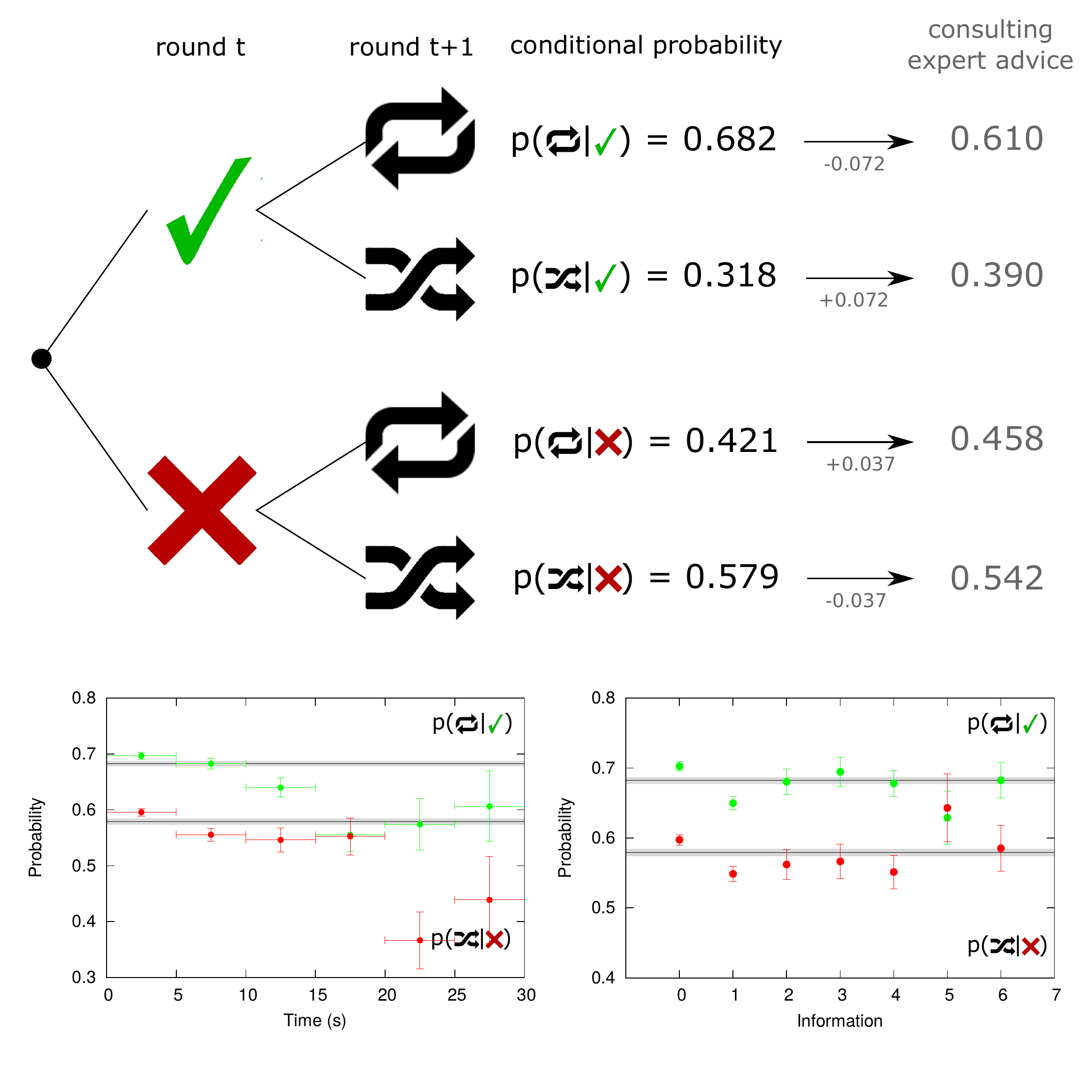}
\caption{\textbf{Decision conditioned to performance: the Win-Stay Lose-Shift emerging strategy}. Empirical conditional probabilities of repeating previous guess are positively correlated with success and failure of the previous guess. The highest probability corresponds to repeating the previous guess when this was correct ($0.682$). The expert's advice partially neutralizes the effect ($0.610$). (down left) The conditional probability decreases when participants spend more time to take the decision in both cases: being ``correct" (green) and being ``wrong" (red) in the previous round. (down right) The conditional probability initially diminishes but when a participant consults more information it oscillates around the mean in both cases: being ``correct" (green) and being ``wrong" (red) in the previous round. Horizontal lines in bottom plots provide aggregated results and shadows their error bars. Error bars represent the standard deviation of a binomial distribution.}
\label{Figure-4}
\end{figure*}

Figure~\ref{Figure-4} focuses on the W-S L-S strategy. In this case, the probability to repeat a successful decision is $0.682$, that is $0.121$ ($19.92$ SD units) higher than the probability to repeat any decision, $0.561$ as shown in the Table~\ref{table1}. In the same way, the probability to change a wrong decision is $0.579$, what is $0.140$ ($21.18$ SD units) greater than the probability to change any decision. Again we observe a behavioural bias being more probable to persist after a successful guess than to change decision after a wrong guess. In this case, shifting the stretegy when guess is ``wrong" can be also related to negative skewness risk \cite{harvey:2000}, to the asymmetric risk (and the increment of market volatility) due to unexpected price drops \cite{masoliver:2006,perello:2003}. 

One obvious question we can formulate is which is the dominant emerging strategy in our experiment. One way to measure the possible differences is by computing the conditional mutual information (see Supplementary Information for the whole analysis). Mutual information between participant's actions and previous market movements conditioned to the previous outcome is $0.05\pm 0.04$ bits, while mutual information between participant's actions and previous outcome conditioned to the previous action market movements $0.07\pm 0.04$ bits. These two conditional information values are telling us that there is non-redundant information. However, market direction (and its subsequent MI strategy) seem to be more relevant since the mutual information conditioned to know the market is higher. 

\begin{figure*}[t]
\centering
\includegraphics[width=0.85\textwidth]{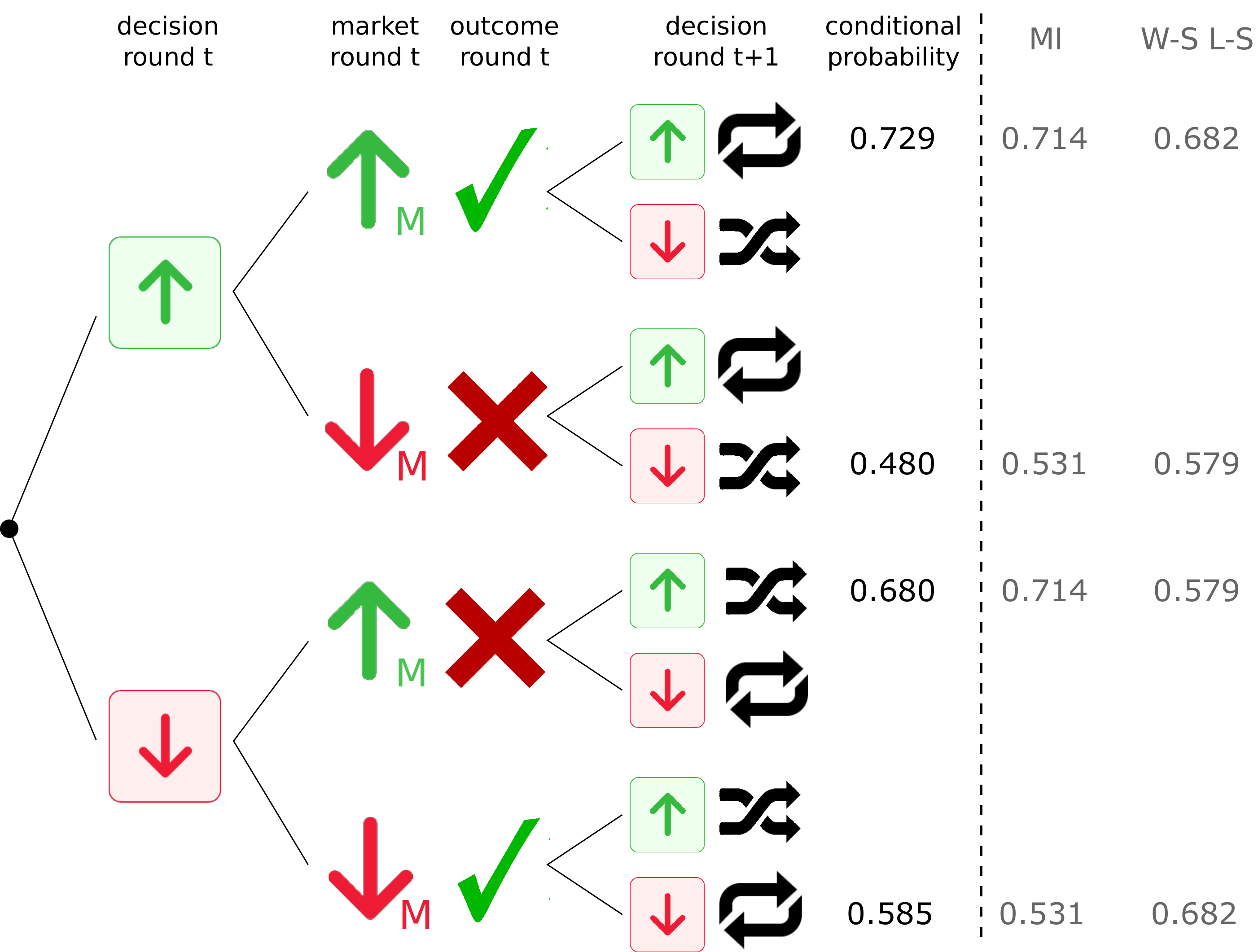}
\caption{\textbf{Two-step Markov chain to observe that Market Imitation is the dominant emerging strategy}. The tree of possibilities for the probability to choose ``up" or ``down" depending on two conditions: the direction of the market in the previous round and the performance of the user (success or not) in the previous round. The tree includes the conditional probabilities of the  Market Imitation and Win-Stay Loss-Shift emerging strategies to observe which of them is the dominant strategy. We observe that Market Imitation conditional probabilities are closer to the aggregated conditional probabilities corresponding to the event to ``Follow and Strategy" than those from the Win-Stay Loss-Shift emerging strategy.}
\label{Figure-5}
\end{figure*}

Another possible approach to evaluate the dominant emerging strategy is the one summarized in Figure~\ref{Figure-5}. In this figure we unfold all possible scenarios with a two-step chain where the MI and W-S L-S strategies appear and can be therefore compared. Thus, for instance, the ``up-success-up" probability is $0.729$ while the probability related to the MI strategy (only conditioned to what market did before, without considering the performance) is $0.714$, thus being much closer than the probability ($0.682)$ related to the W-S L-S strategy. Figure~\ref{Figure-5} shows how MI is systematically closer to the two-step conditional probabilities than the W-S L-S strategy. Therefore, the analysis suggests that the impact of MI strategy is greater than that of the W-S L-S strategy and reinforces the conditional mutual information results.

\subsection{Emerging strategies versus information, time and expert advice}

We test if the conditional probabilities of the MI strategy are also influenced by other variables like the time used to take a decision, the amount of information examined or if the participant consulted the expert's advice. As it can be observed in bottom-left of Figure~\ref{Figure-3}, generally the more time spent to take a decision the lower the values of $p(\uparrow \vert \uparrow_{\rm M})$ and $p(\downarrow \vert \downarrow_{\rm M})$ with respect to the reference value. Such probabilities are however above the reference value when the decision is taken without consulting any information and fall below the reference when one extra panel is consulted (bottom-right Figure~\ref{Figure-3}). The expert's advice also affects the values of $p(\uparrow \vert \uparrow_{\rm M})$ and $p(\downarrow \vert \downarrow_{\rm M})$ by reducing in $0.077$ and $0.017$ respectively. W-S L-S strategy is also susceptible to either time, amount of information and expert advice influences in a similar manner. Regarding the expert's advice, the probability to repeat after a success having consulted the expert is reduced by $0.072$ ($-6.80$ SD units) with respect to the reference level, while the case to chose ``down" when market has fallen and having consulted the expert is $0.037$ ($-2.92$ SD units) below the reference.

We next perform a coarse-grained approach where we tag all the participant's decisions using the two possible labels: if they ``Follow (emerging) Strategy" or if they do ``Not Follow (emerging) Strategy". Thus, ``Follow an Strategy" in the MI case would mean to choose ``up" after market goes up, and choosing ``down" after market goes down (see Figure~\ref{Figure-3}). The values of the other two branches in the tree diagram of Figure~\ref{Figure-3} would be aggregated to conform the ``Not Follow Strategy" probability. We can proceed in the same way with W-S L-S, where ``repeat" after success or ``change" after a wrong guess would mean ``Follow Strategy". Interestingly, we show in the Supplementary Information that for our binomial scheme the aggregation process for the conditional probabilities of the MI and W-S L-S strategies lead to the same events for ``Follow Strategy" and ``Not Follow Strategy".

\begin{figure*}
\centering
\includegraphics[width=1\textwidth]{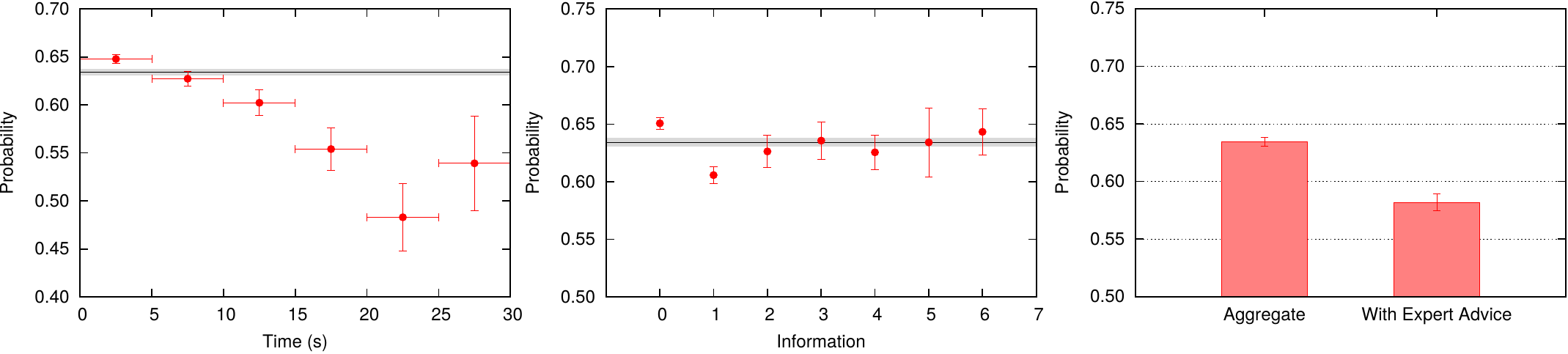}
\caption{\textbf{Aggregated strategies: time, information and expert advice dependences}. (Left) The probability to follow any of the described strategies depending on the time spent to make the decision. The range between 0 to 30 seconds has been divided in bins of 5 seconds. (Center) The probability to follow any of the described strategies depending on the number of different information panels consulted. (Right) The bar on the left accounts for the probability to follow any of the described strategies, whereas the bar on the right indicates the same, but conditioned to having clicked on the expert's advice panel. Solid black line denotes the total probability while the limit of the shaded area and the error bars denote the standard deviation.}
\label{Figure-6}
\end{figure*}

The probability to follow any of the two emerging strategies is $0.634 \pm 0.004$, that is $0.134$ ($25.70$ SD units) over the $0.5$ reference if decisions were random. The probability to ``Follow Strategy" is also affected by time, information and expert's advice as shown in Figure~\ref{Figure-6}, but now the influence seems more evident than in the cases where the two strategies are treated separately. It appears very clear that the tendency to follow any strategy decays with the time spent to take a decision. Moreover, participants without extra information are going to follow much more the emerging strategies in clear contrast to the case with just one extra piece of information (center Figure~\ref{Figure-6}). Surprisingly, more information does not motivate participants to abandon these intuitive strategies since, after consulting more than 1 panel, the probability to follow the strategy increases and stays quite stable close to the same value as to the case of having not consulted extra-information. One possible explanation for this is that two or more different pieces of information are also two different kind of stimulus and they may induce to two different responses (perhaps contradictory responses). Due to an information {\it overload} \cite{bouchaud:2016}, actions would be mostly again governed by emerging strategies. The influence of the expert's advice then becomes very evident in the right panel of Figure~\ref{Figure-6}. We observe that the probability to follow any strategy after the expert is consulted is $0.582$, which is $0.053$ below the reference value ($6.47$ SD units).

Indeed, the simplicity of the coarse-grain analysis make possible to perform a cohort analysis. Table SI.3 of the Supplementary Information shows that all cohort groups tend to follow these intuitive strategies with a probability between $0.6$ and $0.7$, thus providing more universality in this finding. However, there are two exceptions. Firstly, women are more likely to follow these intuitive strategies than men, which is coherent with the fact that they use less time to take decisions and consult less information (including the expert advice) than men. And secondly, children from 0 to 15 years old (corresponding to Primary School) tend to follow the intuitive strategies with a higher intensity than the rest of the groups. Likewise the women group, this is also consistent with the fact that kids take faster and less informed decisions.

\section{Discussion}

%Moreover, stock markets are indeed nowadays dominated by algorithmic trading. Robots are always one step ahead of human decisions since nowadays placing orders into the market just takes few milliseconds~\cite{lewis:2014}.

Facing environments with high levels of uncertainty is a very complicated task for human beings \cite{hommes:2013}. We have serious difficulties when dealing with randomness, we tend to see patterns when there no exist at all~\cite{neuringer:1986}. Moreover, decision-making process involves multiple factors, which may be far from a rational behavior, like stress and panic. Therefore, models that only depart from {\it rationality} and {\it self-interest} could also incorporate concepts linked to how humans cope with uncertain environments~\cite{rabin:1998}.

In this experiment, carried out inside the Pop-Up Experiments framework~\cite{sagarra:2015}, we put a group of volunteers non-expert in finance under such uncertain environment. We asked participants to predict the day-by-day evolution of a series of real historic prices of a certain index, allowing them to consult some information while registering every action \cite{hommes:2013,hommes:2007}.  The 18,436 decisions taken by 283 subjects are very far from being random. Looking at the data we most importantly find two behavioral biases through consistent actions: a preference to guess that the market will move in the same direction as the previous day and a tendency to repeat the previous decisions when they are correct. We identify {\it Market Imitation} and {\it Win-Stay Lose-Shift} strategies as the mechanisms responsible of such biases. These strategies also appear in other contexts~\cite{nowak:1993,jahanshahi:2006}. The coarse-grain approach allows us to identify these strategies as something intuitive because of three reasons: (1) the less time used to make a decision the more likely to follow any of this strategies, (2) the probability to follow any strategy is significantly higher when only a price chart has been consulted and (3) consulting the expert (as a clear exogenous signal) significantly mitigates the likelihood to follow this strategies. The wide range of demographics in our sample allows also to identify that women and children are more intuitive and likely to follow these strategies since in average they take decisions faster and consulting less information than the rest. Moreover, we repeated the same experiment in a conference with different demographics and we confirmed the robustness of our findings. We have also looked at the influence of the previous two steps and we find that, although mutual information values are not relevant anymore, one can still find some traces of behavioural biases. See Supplementary Information for further details on these two last results.

The direct implications of this study point to market policy, traders and market modelling. The pretended advantages of the vertiginous pace of markets and particularly the high-frequency trading are nowadays questioned~\cite{lewis:2014}. In the scale of non-expert individuals, who sometimes might decide to manage their own portfolio, we have seen that fast and uniformed decisions tend to intuitive and pre-established behaviors in contrast with rational and deliberated decisions. However, it should be investigated whether such intuitive behaviours take place also in real markets but there are already studies finding some evidences in traders and fund managers behaviors \cite{hommes:2013,bouchaud:2016}. Our findings anyway supports the idea of decelerating the vertiginous velocity of markets in order to gain rationality and information filtering. From another perspective, our results are of interest for better handling clients expectations of trading companies, avoiding behavioural anomalies in financial analysts decisions and improving the trading digital interfaces where information is set down. It should finally carefully be analyzed the information provided and how it is hierarchized to improve market in many senses \cite{bouchaud:2016}. In a more general way, our results could also help to develop new agent based modelling or stochastic price dynamics to better understand financial bubbles or the effects of asymmetric risk perception to price drops \cite{hommes:2013,hommes:2013b,zhou:2007,perello:2003,masoliver:2006}. The study of the different behavioural biases arising form the emerging strategies can provide some explanation of financial bubbles (Market Imitation, specially for the upwards trends) and how price drops increase market volatility (Win-Stay Loss-Shift, when changing previous decision after a ``wrong" and thus increment market uncertainty). Individuals, as agents who make decisions living in a society impregnated of contingencies impossible to evaluate and constantly updated where we have to take uninformed decisions, must be aware of these intuitive strategies as a fallacies to avoid or lighthouses that help us to sail in the middle of an ocean of uncertainty.

\section{Supporting Information}

% Include only the SI item label in the subsection heading. Use the \nameref{label} command to cite SI items in the text.
\subsection{S1 File}
\label{S1_File}
{\bf Supporting information file.} In the S1 File (PDF) we present further details about the experiment, statistical tests and additional information. (PDF)

\section{Acknowledgments}
We would like to acknowledge the 283 anonymous volunteers who have made this research possible. We are indebted to Barcelona Lab programme through the Citizen Science Office promoted by the Direction of Creativity and Innovation from the Institute of Culture of the Barcelona City Council led by I Garriga for their help and support for setting up the experiment at the DAU Barcelona Festival at Fabra i Coats. We specially want to thank I Bonhoure, N Fern\'andez and P Lorente for all the logistics to make the experiment possible and to O Comas (director of the DAU) for giving us this opportunity. 
In particular, we would like to thank S Saavedra who highly contributed with the original idea for the game and helped us with very useful discussions and comments on the paper. We also thank J Vicens for helping us with the experiment at CAPS2015 and A S\'anchez for a careful reading of the manuscript. 
This work was partially supported by MINECO (Spain) through grants FIS2013-47532-C3-1-P (JD), FIS2013-47532-C3-2-P (JP); by Generalitat de Catalunya (Spain) through Complexity Lab Barcelona (contract no. 2014 SGR 608, JP and MGR) and through Secretaria d'Universitats i Recerca (contract no. 2013 DI 49, JD); by Fundaci\'on Espa\~{n}ola para la Ciencia y la Tecnolog\'ia (FECYT) through the Barcelona Citizen Science Office project of the Barcelona Lab programme; and by the EU through FET-Proactive Project MULTIPLEX (contract no.~317532, JD).

\section{Authors contribution} 
JP, JD and MGR conceived the original idea for the experiment and contributed to the final experimental setup;
JD wrote the software interface for the experiment;
JP, JD and MGR carried out the experiment;
MGR and CS analyzed the data;
JP, JD, CS and MGR discussed the analysis results;
JP, JD, CS and MGR wrote the paper.

\begin{acknowledgments}
Financial support from Direcci\'on General de Investigaci\'on under contract FIS2009-09689 is acknowledged.
\end{acknowledgments}

\end{document}